\documentclass[12pt]{article}
\usepackage{amsmath,amssymb,amscd}
\usepackage{epsfig}
\begin{document}

\begin{center}
{\bf MARKET MILL DEPENDENCE PATTERN IN THE STOCK MARKET: ASYMMETRY STRUCTURE, NONLINEAR CORRELATIONS AND PREDICTABILITY}
\end{center}

\bigskip

\begin{center}
\bf{ \large Andrei Leonidov$^{(a,b,c)}$\footnote{Corresponding author. E-mail leonidov@lpi.ru}, Vladimir Trainin$^{(b)}$, \\Alexander
Zaitsev$^{(b)}$, Sergey Zaitsev$^{(b)}$}
\end{center}
\medskip
(a) {\it Theoretical Physics Department, P.N.~Lebedev Physics Institute,\\
    Moscow, Russia}

(b) {\it Letra Group, LLC, Boston, Massachusetts, USA}

(c) {\it Institute of Theoretical and Experimental Physics, Moscow, Russia}

\bigskip

\bigskip

\bigskip

\begin{center}
{\bf Abstract}
\end{center}

An empirical study of the bivariate probability distribution characterizing a full set of two consecutive price increments $x$ and $y$ for a group of
stocks at time scales ranging from one minute to thirty minutes reveals asymmetric structures with respect to the axes $y=0, \, y=x, \, x=0$ and
$y=-x$. All four asymmetry patterns remarkably resemble a four-blade mill called market mill pattern. The four market mill patterns characterize
different aspects of interdependence between past (push) and future (response) price increments. When analyzed in appropriate coordinates, each
pattern corresponds to a particular nonlinear dependence between the push and the conditional mean of response. Qualitative interpretation of each
pattern is discussed. The market mill pattern is an evidence of complex dependence properties relating past and future price increments and resulting
in various types of nonlinear correlation and predictability.

\newpage

\section{Introduction}

The apparent randomness of stock price dynamics explains a dominant role of the random walk paradigm in theoretical finance. The evolution of stock
prices is indeed close to being random, but a careful statistical analysis reveals a number of characteristic dependence patterns such as
correlations between simultaneous increments of different stocks, statistically significant autocorrelations at intraday timescales, volatility
clustering, leverage effect, etc., that became a part of a standard toolkit for model - building \cite{Man,Lo,SH,BP}. Each of these effects
corresponds to some sort of probabilistic dependence between lagged and/or simultaneous price increments.

A general way of quantifying the probabilistic dependence providing a full description of the set of quantities under consideration is to specify the
corresponding multidimensional probability distribution, from which one can calculate all quantities of interest such as various conditional
distributions and their moments. Generically such distributions can depend on simultaneous and lagged price increments, traded volumes, etc.

Our approach to studying the dynamical patterns in the stock price evolution \cite{LTZ05,LTZZ05a,LTZZ05b} is based on a direct analysis of such
multidimensional probability distributions. A simplest case we concentrate upon in our studies \cite{LTZ05,LTZZ05a,LTZZ05b} is that of a bivariate
distribution describing the interdependence of two price increments in two coinciding or non-overlapping time intervals. Even here there is quite a
few cases to analyze: the increments can be simultaneous and belong to different stocks or lagged and belong to the same stock or different stocks,
the time lag separating the intervals can vary, etc. The interdependence between lagged price increments for the same stock was termed "horizontal"
and the one between simultaneous price increments of different stocks termed "vertical" in \cite{LTZ05}, where both horizontal and vertical
dependencies were considered. In the present paper we focus on the horizontal case only.

Despite of the fundamental importance of the issue, a list of papers on the multidimensional probability distributions of stock price increments is,
to the best of our knowledge, not abundant. A theoretical analysis of bivariate distribution of returns in two consecutive intervals in the
particular case of Levy-type marginals was performed by Mandelbrot \cite{M63}, where some interesting geometric features of this distribution both
for the case of independent and dependent returns were described. As discussed in \cite{MB}, the bivariate distribution in question can be considered
as a "fingerprint" reflecting a nature of the pattern embracing the two consecutive returns.

Recently an analysis of conditional dynamics of volatility exploiting conditional distributions was described in \cite{CJY05}. Let us also mention
direct studies of the above-described bivariate distributions for returns in consequent time intervals that revealed interesting phenomena such as
the "compass rose"  \cite{CL96,AV04,V04}. The conclusions on whether the discovered patterns are related to predictability differed: whereas
\cite{CL96} found no predictability, in \cite{V04} an interesting pattern of wavelet-filtered series corresponding to partial predictability was
found. The first moment of corresponding conditional distribution for daily time intervals was studied in \cite{BM03}. Another line of research is an
explicit analytical reconstruction of the bivariate distribution in question using copulas, see e.g. \cite{EMS99,ELM01,MS01,JR01}.

Any dependence between lagged price increments means that certain characteristics of price evolution are predictable. The issues of predictability
and market efficiency are of paramount importance for theoretical finance. The analysis of the present paper is relevant in this context because a
key quantity characterizing predictability, the conditional mean  \cite{F70,F76a,F76b,F91,L89}, is determined by the conditional distribution of the
increment in the second interval at given increment in the first one.  Let us emphasize that the profitability of a strategy given the existence of
the nonzero conditional mean requires, in each and every case, a separate analysis. A particular example of potential profit from high frequency
anticorrelations of stock price returns was discussed in \cite{BP}. The properties of conditional mean are most often discussed in the context of
verifying the martingale property of the financial series \cite{M66} which is, in turn, a generally accepted outcome of market efficiency.

The paper is organized as follows.

In paragraph 2.1 we describe the market data used in our study and elaborate on basic points behind the probabilistic methodology used in its
analysis.

Main empirical results of the paper are described in paragraph 2.2. We begin with discussing the symmetry properties of the bivariate distribution of
price increments in two consecutive time intervals, termed push and response. The distribution is shown to be approximately invariant with respect to
rotations at a multiple of $\pi/2$. This gives rise to a number of relations between the probabilities of certain two-step processes. We continue
with discussing four asymmetries of the bivariate distribution with respect to the axes $y=0$, $y=x$, $x=0$ and $y=-x$: asymmetry of conditional
response, asymmetry with respect to increment's ordering , asymmetry of conditional push and asymmetry of mirror orderings correspondingly.
Remarkably all of these asymmetries are characterized by the same geometric pattern termed market mill. The asymmetries are shown to result from the
basic geometry of the full distribution described at the beginning of the section.

In paragraph 2.3 we discuss major outcomes from the market mill patterns. In particular a nonlinear dependence of the conditional mean response on
the push clearly demonstrates the nonlinear nature of correlations between the two variables in the considered bivariate probability distribution.
The market mill pattern signals the market inefficiency at intraday time scales resulting from temporal interdependencies between consecutive price
increments.

We conclude by formulating the main results and outline of the future work in paragraph 3.

\section{Asymmetry patterns in price dynamics}

\subsection{Data and Methodology}

Our study of high frequency dynamics of stock prices are based on data on the prices of 100 stocks traded in NYSE and NASDAQ in 2003-2004 sampled at
1 minute frequency\footnote{All holidays, weekends and off-market hours are excluded from the dataset. The list of stocks is given in the Appendix.}.

A method of analyzing dependence patterns in financial market dynamics adopted in the present paper, see also \cite{LTZ05,LTZZ05a,LTZZ05b}, is based
on a direct examination of multivariate probability distributions and conditional distributions constructed from them. Below we shall restrict our
consideration to the bivariate distribution of stock price increments in two non-overlapping time intervals.

Let us consider two non-overlapping time intervals of length $\Delta T_1$ and $\Delta T_2$. The time ordering is such that the interval $\Delta T_2$
follows after the interval $\Delta T_1$. Generically the end of the first interval and the beginning of the second can be separated by some time
interval but in most of cases under consideration the second interval immediately follows the first one. In this case we are dealing with two price
increments $\delta p (\Delta T_1) = p(t)-p(t-\Delta T_1)$ and $\delta p (\Delta T_2) = p(t+\Delta T_2)-p(t)$. In what follows we shall denote the
price increment in the first interval $\delta p (\Delta T_1)$ by $x$ (push) and that in the second one $\delta p (\Delta T_2)$ by $y$ (response).
Below we consider the case of $\Delta T_1=\Delta T_2=\Delta T$ and study four time scales $\Delta T=$ 1 min, 3 min., 6 min. and 30 min.

The full probabilistic picture of interrelation between the push and response is given by a bivariate probability density  ${\cal P} (x,y)$. This is
the most comprehensive quantity from which other quantities of practical interest could be derived. In particular, one can compute the conditional
probability distribution ${\cal P}(y|\,x)={\cal P}(x,y)/{\cal P}(x)$ describing the probabilistic shape of response at given push.

Let us stress that in the present paper we consider the probability distribution characterizing a full set of price increment pairs for a group of
stocks so that a set of events (pairs of increments) under consideration unifies all subsets of events characterizing individual stocks. Studying the
properties of thus obtained bivariate distribution ${\cal P}(x,y)$ is the main goal of the present paper and the companion paper \cite{LTZZ05a}.
Bivariate distributions for individual stocks are considered in the companion paper \cite{LTZZ05b}.

Let us now comment on our choice of price increments and not price returns as basic variables. The question of a choice of the "right" variable is by
no means trivial. The full picture may well be the one using both increments and returns at different time scales. At intraday time scales the market
participants use strategies based on evaluating price moves in dollars, while at larger time scales (e.g. weeks) they are rather formulated and
realized in terms of returns. The particular scheme interpolating between price increments at small time scales and returns at large ones is
described in \cite{BP}. Let us emphasize that we have checked that the main results described in the paper are valid for normalized returns as well.

\subsection{Empirical results: market mill patterns}

Let us now turn to discussing the structure of push - response relationship by considering the market data on price increments in two consecutive 1 -
minute intervals. As discussed in the previous section, the full probabilistic picture of the two-step push - response process is described by the
probability distribution ${\cal P}(x,y)$. The two - dimensional projection of $\log_8 {\cal P}(x,y)$ \footnote{In all plots we are using the
$\log_{2^n}$ scale for the probability distribution so that the equiprobability lines differ by the factor of $2^n$. The particular values of $n$ are
chosen to show the distribution geometry in the most clear fashion. The borders between different colors correspond to the equiprobability lines.} is
plotted in Fig.~10 \footnote{The paper contains charts in "eps" and "png" formats. The numeration is such that the "png" charts follow the "eps"
ones.}. This distribution turns out to have many remarkable properties that we analyze below and in the companion papers \cite{LTZZ05a,LTZZ05b}.

To facilitate the analysis let us sketch the shape of the equiprobability levels in Fig.~1,
\begin{figure}[h]
 \begin{center}
 \leavevmode
 \epsfysize=7cm
 \epsfbox{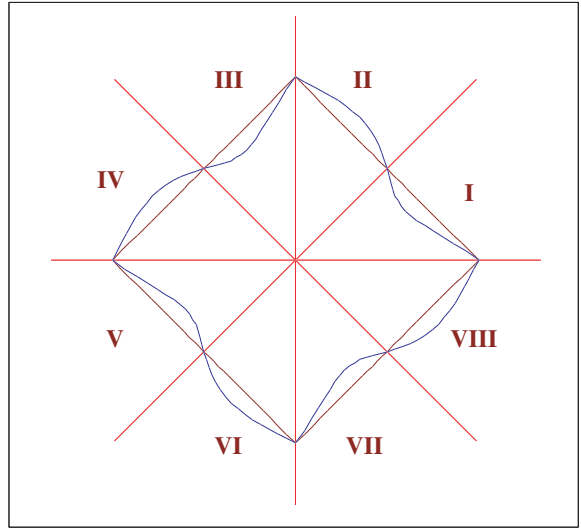}
 \end{center}
\caption{Sketch of the equiprobability levels of the bivariate distribution ${\cal P} (x,y)$. Straight lines are drawn to guide the eye.}
\end{figure}
in which the $xy$ plane is divided into sectors numbered counterclockwise from I to VIII.

The first striking property of the bivariate distribution ${\cal P}(x,y)$ is its approximate invariance with respect to rotations at the multiples of
$\pi/2$. In terms of sample paths composed by the increments $\pm \zeta_1$ and $\pm \zeta_2$ the symmetry with respect to rotations at multiples of
$\pi/2$ leads to a chain of equalities\footnote{A quantitative analysis of these properties is presented in the companion paper \cite{LTZZ05a}.}
\begin{equation}
{\cal P}(\zeta_1,\zeta_2) = {\cal P}(-\zeta_2,\zeta_1) = {\cal P}(-\zeta_1, - \zeta_2) = {\cal P}(\zeta_2,-\zeta_1)
\end{equation}
Second, let us stress that naively expected commutativity of price increments under addition ${\cal P}(\zeta_1,\zeta_2)= {\cal P}(\zeta_2,\zeta_1)$
is violated by the asymmetries of the bivariate distribution ${\cal P}(x,y)$  analyzed in full details in the next paragraph. Thus  a simple
interchange of price increments $\zeta_1 \leftrightarrow \zeta_2$ {\it is not a symmetry} of price dynamics.

The third crucial property of the ${\cal P}(x,y)$ is that all even sectors are "stronger" (containing more probability) than the odd ones. This leads
to a number of remarkable consequences described in the next paragraph.

\subsubsection{Asymmetry of conditional response}

Let us first analyze the symmetry of the distribution ${\cal P}(x,y)$ with respect to the axis $y=0$. In term of sectors shown in Fig.~1 we compare
sector I versus sector VIII, sector II versus sector VII, etc. To arrange a convenient framework for analyzing this question let us separate the
two-dimensional distribution ${\cal P} (x,y)$ into a sum of symmetric ${\cal P}^s_I$\footnote{We use a subscript "I" to index the first of the four
asymmetries considered in the paper.}  and antisymmetric ${\cal P}^a_I$ components:
\begin{equation}\label{sepdis1}
 {\cal P} (x,y) \, = \, {\cal P}^s_I + {\cal P}^a_I \,
\end{equation}
where
\begin{equation}\label{sepdis11}
{\cal P}^s_I \, = \, \frac{1}{2} \left ( {\cal P} (x,y) + {\cal P}(x,-y)  \right ) \,\,\,\,\, {\rm and} \,\,\,\,\, {\cal P}^a_I \, = \, \frac{1}{2}
\left ( {\cal P} (x,y) - {\cal P}(x,-y)  \right )
\end{equation}
This decomposition is of course not a sum of two probability densities. To provide a probabilistic meaning for the asymmetric term $ {\cal P}^a_I
(y,x)$ one should consider only its positive part.

To get a better visual picture of the asymmetric component  ${\cal P}_I^a(x,y)$ let us consider its positive part ${\cal P}_I^{a (p)}(x,y) = \theta
\left[ {\cal P}_I^{a}(x,y) \right] \, {\cal P}_I^{a}(x,y)$, where $\theta [ \cdot ]$ is a Heaviside step function. The definition of  ${\cal P}_I^{a
(p)}(x,y)$ guarantees that no information is lost when imposing this restriction. In Fig.~8 we show a two-dimensional projection of the surface
$\log_4 {\cal P}_I^{a (p)}(x,y)$ in the range $  \$ \, -0.3  <= x,y <= \$ \, 0.3 $  for the case of two consequent 1 - minute intervals.

We see that the plot in Fig.~8 demonstrates a remarkable four - blade mill - like pattern or simply the market mill pattern. Before analyzing this
asymmetry in more details, let us point out that the overall asymmetry, if measured with respect to the symmetric component, is of order of 5
percent, so we are in fact dealing with a small asymmetric component on top of the dominating symmetric one.

To study the structure of the response asymmetry ${\cal P}_I^{a (p)}$ in more details, let us consider the sections of its surface by the constant
push planes $x={\rm const.}$ In Fig.~9 we show three response profiles for $\log_4 {\cal P}_I^{a (p)}$ for the push values of $x=\$ \,\,
0.01,\,0.07,\,0.25 \, $.

For small pushes we observe a clear example of the skewed distribution. The small negative response after the small initial positive push happens
quite often, but is outweighted by the positive tail corresponding to big positive response. So, a reaction to the small push is either the small
contrarian response or the big followup in the direction of the initial push separated by the gap. At medium pushes the situation becomes more
balanced: contrarian mechanism eats away a large chunk of the response asymmetry space. Finally, for large pushes the response is purely contrarian.

A dependence of the response pattern on the magnitude of the push is best quantified by the conditional distribution  $ {\cal P} (y|\,x) = {\cal P}
(x,y)/{\cal P}(x)$. In particular the market mill structure shown in Fig.~8 leads to a very nontrivial z-shaped nonlinear behavior of the conditional
mean response $\langle y \rangle_x \, = \, \int dy \, y \, {\cal P} (y|\,x)$. To illustrate this phenomenon we plot in Fig.~2 the dependence of the
conditional mean response on the push at three time scales $\Delta T=$ 1 min., 3 min. and 6 min.
\begin{figure}[h]
 \begin{center}
 \leavevmode
 \epsfxsize=14cm
 \epsfbox{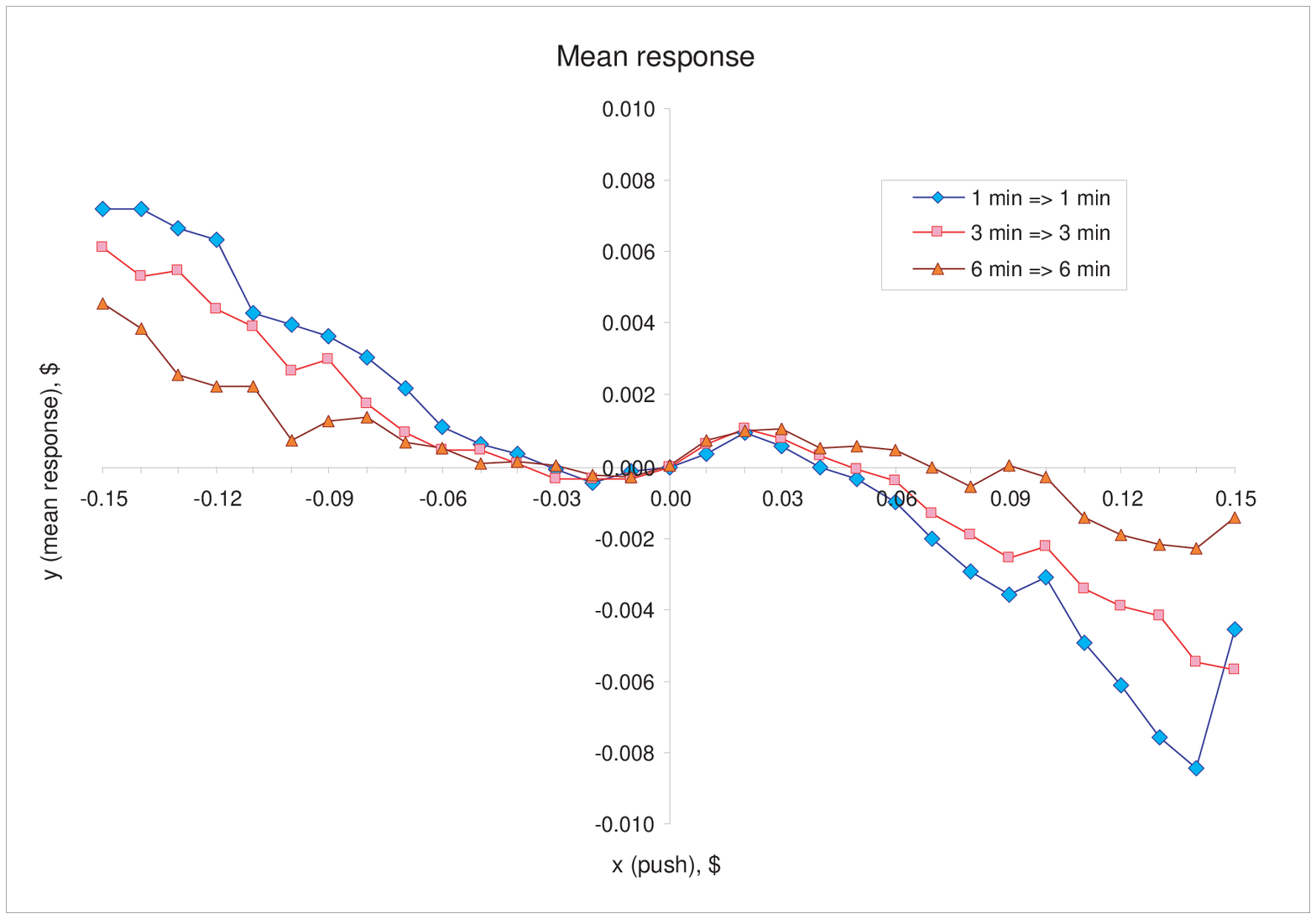}
 \end{center}
  \caption{Mean conditional response, 1 min $\Longrightarrow$ 1 min, 3 min $\Longrightarrow$ 3 min, 6 min $\Longrightarrow$ 6 min}
\end{figure}

To clarify the relation between the market mill pattern shown in Fig.~8 and the behavior of the conditional response let us note that the
decomposition of the two-dimensional distribution in Eqs.~(\ref{sepdis1},\ref{sepdis11}) generates a corresponding decomposition of the conditional
distribution ${\cal P}(y|\,x)$ into symmetric and antisymmetric parts
\begin{equation}\label{sepdis2}
 {\cal P} (y|\,x) \, = \, {\cal P}^s_I (y|\,x) + {\cal P}^a_I (y|\,x)
\end{equation}
It is easy to check that the conditional response $\langle y \rangle_x$ is a functional of the asymmetric component only:
\begin{equation}\label{condmean}
 \langle y \rangle_x \equiv \int dy \, y \, {\cal P} (y|\,x) \, = \,  \int dy \, y \, {\cal P}^a_I (y|\,x)
\end{equation}
so that the nontrivial nonlinear response shown in Fig.~2 directly follows from the market mill pattern of Fig.~8. In terms of sectors shown in
Fig.~1 the nonlinear mean response follows from the fact that at small $x>0$ contribution of sector II is stronger than that of sector VIII while at
larger $x$ sector VIII starts dominating over sector II, etc.

Let us illustrate the asymmetry of conditional mean by considering a sample price trajectory in the $xy$ plane composed by two consecutive price
increments $\zeta_1$ and $\zeta_2$. The existence of the market mill structure shown in Fig.~8 means that the probability ${\cal P}(\zeta_1,\zeta_2)$
of the combination $(\zeta_1,\zeta_2)$ is not equal to that of the combination $(\zeta_1,-\zeta_2)$, see the diagram in Fig.~3.
\begin{figure}[h]
 \begin{center}
 \leavevmode
 \epsfxsize=8cm
 \epsfbox{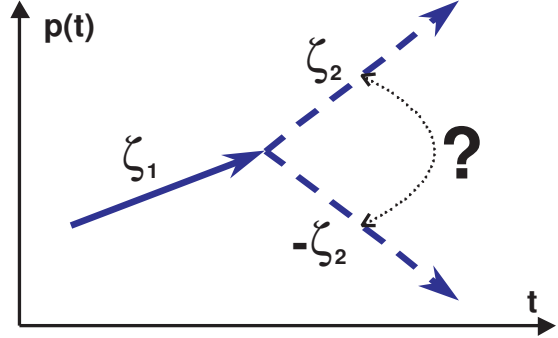}
 \end{center}
 \caption{Diagram illustrating the conditional response asymmetry}
\end{figure}

How does the market mill asymmetry depend on the length of the time interval? To study this question we have examined a set of intervals of length
$\Delta T = 6$ and $30$ minutes. The corresponding asymmetries are shown in Figures.~11-12.
All the plots show the presence of the market mill structure which is progressively becoming more noisy with the growing length of the interval. Let
us also mention that if we have a gap between the two time intervals under consideration, the market mill structure is still present. This is
illustrated in Fig.~13 where we show the asymmetry for two 1 minute intervals separated by 1~minute gap.

\subsubsection{Asymmetry with respect to increment's ordering}

Let us now analyze the symmetry properties of ${\cal P}(x,y)$ with respect to the axis $y=x$. The appropriate decomposition of ${\cal P}(x,y)$ into
symmetric and antisymmetric contributions reads
\begin{equation}\label{sepdis2}
 {\cal P} (x,y) \, = \, {\cal P}^s_{II} + {\cal P}^a_{II} \,
\end{equation}
where
\begin{equation}\label{sepdis21}
{\cal P}^s_{II} \, = \, \frac{1}{2} \left ( {\cal P} (x,y) + {\cal P}(y,x)  \right ) \,\,\,\,\, {\rm and} \,\,\,\,\, {\cal P}^a_{II} \, = \,
\frac{1}{2} \left ( {\cal P} (x,y) - {\cal P}(y,x)  \right )
\end{equation}
A two-dimensional projection of $\log_4 {\cal P}^{a (p)}_{II}$, where ${\cal P}^{a (p)}_{II}$ is a positive component of the asymmetric part ${\cal
P}^a_{II}$, is plotted in Fig.~14 showing the market mill pattern similar to the one for the asymmetry of the conditional response in paragraph
2.2.1. The difference is that for the market mill in paragraph 2.2.1 the first blade (sector 2) with respect to the axis $y=0$ is one sector away
from it, while here the first blade (the same sector 2) is the sector closest to the axis $y=x$. This difference results in specific z-shaped pattern
discussed below, see Fig.~4.

Another way of analyzing the symmetry properties of ${\cal P}(x,y)$ is to introduce new coordinates
\begin{equation}
 z \, = \, \frac{1}{\sqrt{2}} (x+y) \, ; \,\,\,\,\,\,\,\,\, {\bar z} \, = \, \frac{1}{\sqrt{2}} (y-x)
\end{equation}
which means that we have rotated the original coordinate axes at $\phi=\pi/4$. This specific rotation is chosen because geometrically the symmetry
with respect to the transformation $x \longleftrightarrow y$ is the symmetry upon reflection with respect to the axis $x=y$, which in the new
coordinates $(z, {\bar z})$ a symmetry with respect to the axis ${\bar z}=0$. In new coordinates
\begin{equation}\label{sepdispath1}
 {\cal P}^s_{II}  \, = \, \frac{1}{2} \left ( {\cal P} (z,{\bar z}) + {\cal P}(z,-{\bar z})  \right ) \,\,\,\,\ {\rm and} \,\,\,\,\, {\cal P}^a_{II}
  \, = \, \frac{1}{2} \left ( {\cal P} (z,{\bar z}) - {\cal P}(z,-{\bar z})  \right )
\end{equation}
The properties of the distribution (\ref{sepdispath1}) can be conveniently analyzed by considering a conditional distribution ${\cal P} ({\bar z} |\,
z)$. In Fig.~4 we plot the conditional mean $\langle {\bar z} \rangle_z$ versus $z$ at $1$ min. time scale.
\begin{figure}[h]
 \begin{center}
 \leavevmode
 \epsfxsize=14cm
 \epsfbox{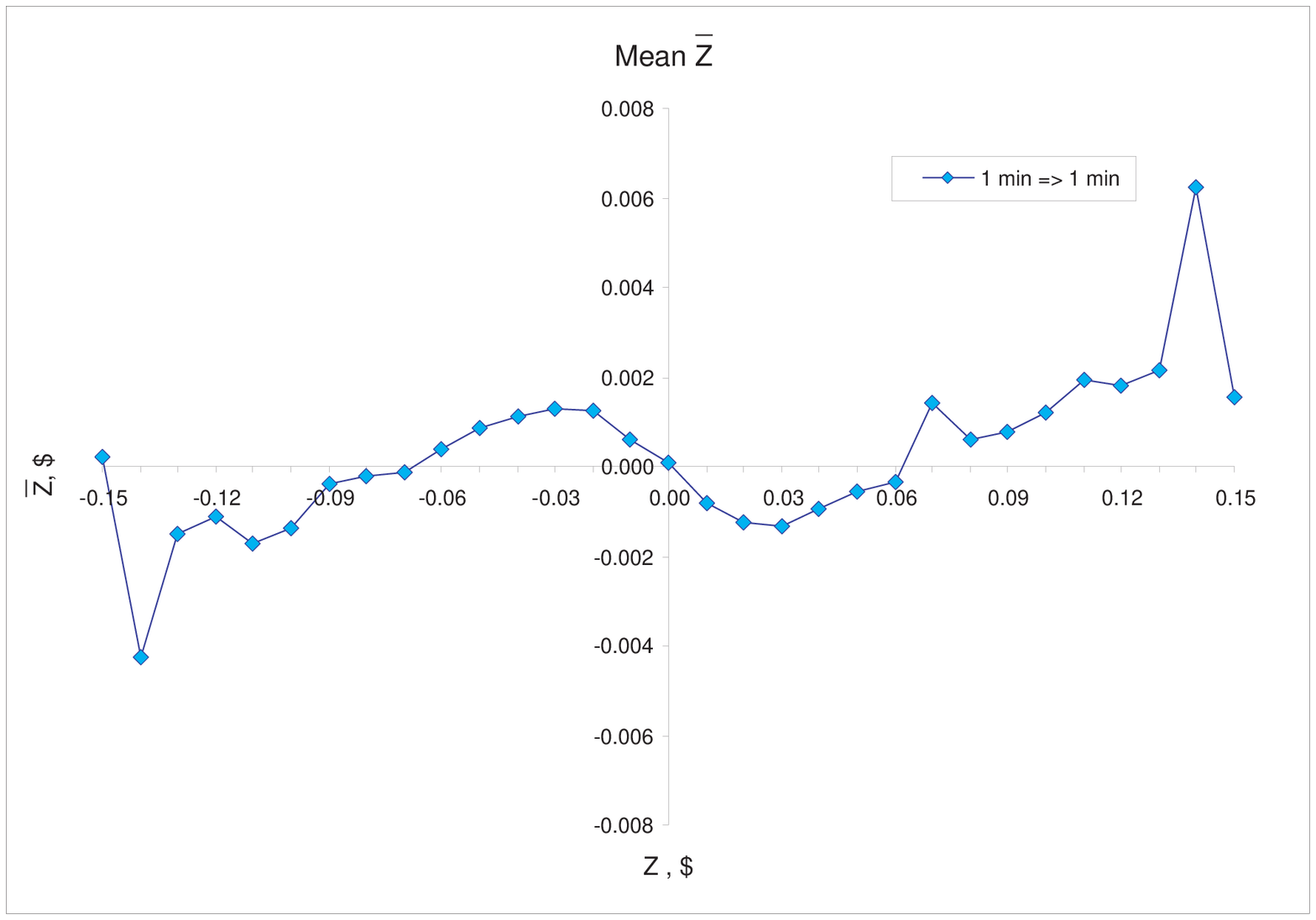}
 \end{center}
  \caption{Conditional mean with respect to the axis $y=x$, 1 min $\Longrightarrow$ 1 min }
\end{figure}
We see the z-shaped structure analogous to the one discussed in the previous paragraph, with the difference that the conditional mean ${\bar z}$ is
now trend-following at large $|z|>z_0$, where $z_0 \sim \$ \, 0.1$, and contrarian at small $|z|<z_0$. This asymmetry of the conditional mean
$\langle {\bar z} \rangle_z$ is, of course, another manifestation of the basic asymmetry with respect to the transformation $x \longleftrightarrow
y$. To illustrate conclusions following from the pattern plotted in Fig.~4, let us again consider the sample path $(\zeta_1,\zeta_2)$ and list the
basic patterns:
\begin{eqnarray}
  \zeta_1+\zeta_2< - z_0 & \Longrightarrow \,\,\, \rm{on \,\, average}& \zeta_2 < \zeta_1 \nonumber \\
  -z_0<\zeta_1+\zeta_2<0 & \Longrightarrow \,\,\, \rm{on \,\, average}& \zeta_2 > \zeta_1 \nonumber \\
  0 <\zeta_1+\zeta_2< z_0 & \Longrightarrow \,\,\, \rm{on \,\, average}& \zeta_2 < \zeta_1 \nonumber \\
  \zeta_1+\zeta_2 > z_0 & \Longrightarrow \,\,\, \rm{on \,\, average}&  \zeta_2 > \zeta_1
\end{eqnarray}

A clear interpretation of the asymmetry with respect to the axis $y=x$ is obtained by comparing the probabilities of two sample paths
\begin{itemize}
\item{{\bf Path 1}:  Push $\zeta_1$, response $\zeta_2$, $\zeta_1+\zeta_2$ fixed, probability ${\cal P}(\zeta_1,\zeta_2)$}
\item{{\bf Path 2}:  Push $\zeta_2$, response $\zeta_1$, $\zeta_2+\zeta_1$ fixed, probability ${\cal P}(\zeta_2,\zeta_1)$}
\end{itemize}
This is illustrated in the diagram in Fig.~5.
\begin{figure}[h]
 \begin{center}
 \leavevmode
 \epsfxsize=8cm
 \epsfbox{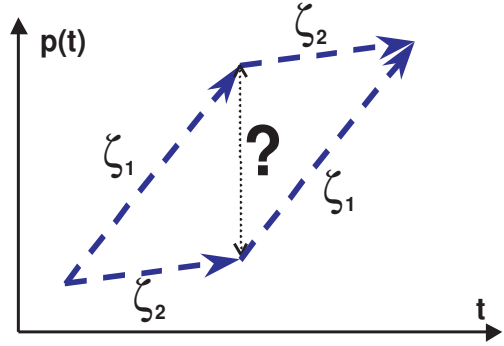}
 \end{center}
\caption{Diagram illustrating the asymmetry with respect to increment's ordering}
\end{figure}
The market mill pattern corresponding to the asymmetry of ${\cal P}(x,y)$ with respect to the axis $y=x$ leads therefore to the striking conclusion
that the ordering of the price increments is probabilistically essential!

\subsubsection{Conditional push asymmetry}

Let us turn to the analysis of the symmetry of the distribution ${\cal P}(x,y)$ with respect to the axis $x=0$. The appropriate separation of the
full distribution into symmetric and antisymmetric parts reads
\begin{equation}\label{sepdisconpush}
 {\cal P} (x,y) \, = \, {\cal P}^s_{III} + {\cal P}^a_{III} \, ,
\end{equation}
where
\begin{equation}\label{sepdisconpush1}
 {\cal P}^s_{III} \, = \, \frac{1}{2} \left ( {\cal P} (x,y) + {\cal P}(-x,y)  \right ) \,\,\,\, {\rm and} \,\,\,\,
 {\cal P}^a_{III} \, = \,  \frac{1}{2} \left ( {\cal P} (x,y) - {\cal P}(-x,y)  \right )
\end{equation}
We are thus dealing  with an "inverse" relation between the push and response, namely, with the structure of push patterns leading to a given
response. A two-dimensional projection of $\log_4 {\cal P}^{a(p)}_{III}$, where  ${\cal P}^{a(p)}_{III}$ is a positive component of the asymmetric
part  ${\cal P}^a_{III}$, is plotted in Fig.~15.  It shows the same market mill pattern as characterizing the asymmetry of conditional response
discussed in paragraph 2.2.1. All the discussion on the asymmetry of conditional response can be repeated here for the conditional push. This
includes, in particular, the zigzag structure of the dependence of mean conditional push on response $\langle x \rangle_y$.

Another way of looking at this problem is a rotation of the axes at $\phi = \pi/2$:
\begin{equation}
z \, = \, y \, ; \,\,\,\,\,\,\,\,\, {\bar z} \, = \, -x
\end{equation}
followed by the corresponding decomposition of the full probability distribution with respect to the axis ${\bar z}=0$.

The asymmetry of conditional push corresponds to considering the probabilities of the two sample paths $(\zeta_1,\zeta_2)$ and $(-\zeta_1,\zeta_2)$,
respectively ${\cal P}(\zeta_1,\zeta_2)$ and ${\cal P}(-\zeta_1,\zeta_2)$, so that we are interested in the properties of the "prehistory" of the
increment in the second interval $\zeta_2$, see the diagram in Fig.~6.
\begin{figure}[h]
 \begin{center}
 \leavevmode
 \epsfxsize=8cm
 \epsfbox{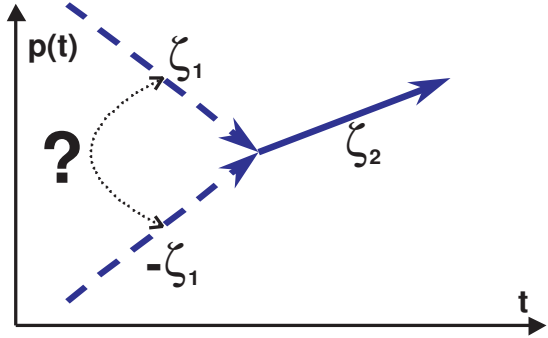}
 \end{center}
\caption{Diagram illustrating the conditional push asymmetry}
\end{figure}

\subsubsection{Asymmetry of mirror orderings}

Let us finally analyze the symmetry properties of ${\cal P}(x,y)$ with respect to the axis $y=-x$. The corresponding symmetric and antisymmetric
components read
\begin{equation}\label{sepdistraj}
 {\cal P} (x,y) \, = \, {\cal P}^s_{IV} + {\cal P}^a_{IV}
\end{equation}
where
\begin{equation}\label{sepdistraj1}
 {\cal P}^s_{IV} \, = \,  \frac{1}{2} \left ( {\cal P} (x,y) + {\cal P}(-y,-x)  \right ) \,\,\,\, {\rm and} \,\,\,\,
 {\cal P}^a_{IV} \, = \, \frac{1}{2} \left ( {\cal P} (x,y) - {\cal P}(-y,-x)  \right )
\end{equation}
A two-dimensional projection of $\log_4 {\cal P}^{a(p)}_{IV}$, where ${\cal P}^{a(p)}_{IV}$ is a the positive component of the asymemtric part ${\cal
P}^a_{IV}$, is plotted for the case of two consecutive 1 - minute intervals  in Fig.~16. We see the same four - blade market mill pattern as in
paragraph 2.2.2.

A useful view on the symmetry with respect to the axis $y=-x$ is provided by using the rotated coordinate frame, which in this case is obtained by
rotation at $\phi=3 \pi/4$:
\begin{equation}
 z \, = \, \frac{1}{\sqrt{2}} (x-y) \, ; \,\,\,\,\,\,\,\,\, {\bar z} \, = \, \frac{1}{\sqrt{2}} (x+y)
\end{equation}
The conditional mean $\langle {\bar z} \rangle_z$ has the same shape as in the case of the asymmetry with respect to the increment's ordering, but
now this is a conditional mean of the total increment $z_1+z_2$ at given difference $z_1-z_2$.

To elucidate the meaning of the asymmetry with respect to the axis $y=-x$, let us consider the probabilities of two sample paths
\begin{itemize}
\item{{\bf Path 1}:  Push $\zeta_1$, response $\zeta_2$, $\zeta_1+\zeta_2=z^*$}, probability ${\cal P}(\zeta_1,\zeta_2)$
\item{{\bf Path 2}:  Push $-\zeta_2$, response $-\zeta_1$, $-\zeta_2-\zeta_1=-z^*$}, probability ${\cal P}(-\zeta_2,-\zeta_1)$
\end{itemize}
This is illustrated by the diagram in Fig.~7.
\begin{figure}[h]
 \begin{center}
 \leavevmode
 \epsfxsize=8cm
 \epsfbox{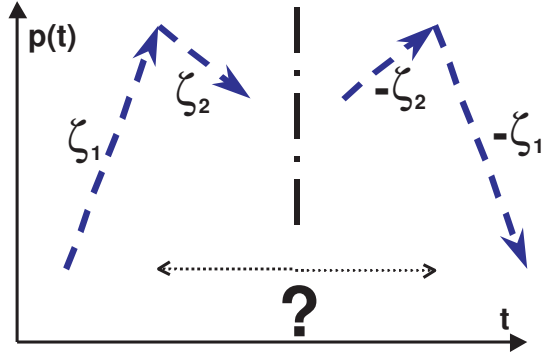}
 \end{center}
\caption{Diagram illustrating the asymmetry of mirror orderings}
\end{figure}
Note that the second path is a mirror one with respect to the first. One could also describe the Path 2 as a trajectory reversal. The meaning of the
asymmetry with respect to the axis $y=-x$ is thus that the probabilities associated with two mirror trajectories are not equal.

\subsection{Discussion}

Let us add a few more points to the discussion of the properties of the structure of probability distribution in the push - response plane.

First, the existence of the z-shaped conditional mean response implies the existence of a nonlinear predictability\footnote{By nonlinear
predictability we mean the nonlinear dependence of conditional mean response on the push.}. Does this predictability reflect market inefficiency at
the timescales under consideration? The generic probabilistic test of the market efficiency is model-dependent \cite{F70,F76a,F76b,F91,L89}. A
simple, although not totally generic picture of the efficient market is that of a martingale dynamics \cite{M66}. In this picture the full
implementation of available information into current stock price means that the expected value of this price in the future equals its current one,
which in the framework of this paper means that $\langle y \rangle_x = 0$ for all $x$. The nonzero conditional mean is thus explicitly breaking the
martingale condition $\langle y \rangle_x = 0$, see Eq.~(\ref{condmean}), and thus, within the generally accepted paradigm, signals market
inefficiency.

Second, it is of substantial interest to view our results on the conditional response in connection with an usual measure of dependence between
stochastic variables - the covariance. The unconditional mean increments in all intervals considered are so small that one can safely neglect it. The
covariance is than equal to
\begin{equation}\label{conmeanresp}
 \langle x\,y \rangle = \int dx\,x\,P(x)\,\langle y \rangle_x
\end{equation}
It is clear the nonlinear structure of the conditional response $\langle y \rangle_x$ strongly affects the value of conditional mean. For example, if
the conditional response $\langle y \rangle_x$ is even in the push, say $\langle y \rangle_x \propto x^2$, the covariance and thus the linear
correlation coefficient are, for symmetric marginal distribution $P(x)$, identically zero. A somewhat more realistic example of $\langle y \rangle_x
= a x - bx^3$ gives $\langle x\,y\rangle = a \langle x^2 \rangle - b \langle x^4 \rangle$, so that in this case the value of the linear correlation
coefficient depends, through parametrization of the conditional response, on higher moments of the marginal distributions $P(x)$. We see that
generically the linear correlation does not provide an adequate description of the response. In particular, the zigzag form of the conditional
response makes description in terms of linear correlation unreliable. Let us notice that even the sign of the high frequency autocorrelation does not
seem to be well established. The popular view is that at short timescales the increments (returns) are anticorrelated \cite{BP}. A detailed analysis
of high frequency returns in \cite{LGCMPS99} showed, however, a positive autocorrelation. This ambiguity can well be related to the nonlinear
structure of the conditional response discussed in the present paper.

Third, the existence of market mill structure means that price evolution is characterized by an explicit breakdown of the symmetry with respect to
time reversal, which in the considered case is a symmetry with respect to the push-response interchange $ x \leftrightarrow y$. Previously the time
reversal asymmetry effects were found only in the context of volatility dynamics \cite{LZ03,LB05}. Let us note, that the breakdown of the time
reversal symmetry and the zigzag structure of the mean response provides valuable information on the two-dimensional probability distribution ${\cal
P}(x,y)$. For the popular elliptical distributions, i.e. those dependent on the quadratic form dependent on $x$ and $y$, the mean response is always
linear in the push and the distribution function is symmetric with respect to $x \leftrightarrow y$ transformation. Therefore, the two-dimensional
distribution of consecutive increments is, in fact, non-elliptical, which is a serious obstacle in developing an analytical description of this
distribution. Let us mention that in \cite{BDE03} it was shown that the bivariate distributions of the returns at FX market is not elliptical at
short intraday timescales.

The above-described remarkable market structures characterize the bivariate distribution for the whole ensemble of price increments of all the stocks
considered. Does this blend also give an adequate description for asymmetry characterizing individual stocks? Careful analysis shows that individual
asymmetry patterns vary strongly from security to security. This makes the stability in these individual patterns even more unexpected. One could say
that each stock is characterized by its own "fingerprint" of the asymmetry of push-response distribution. A detailed analysis of this issue will
appear in the forthcoming work \cite{LTZZ05b}. Here we just show in Fig.~17 three individual characteristic patterns corresponding to the cases with
the dominance of nonlinear response, ordinary correlation and anticorrelation patterns respectively. To illustrate the stability of the asymmetry
patterns in Fig.~17 they are shown in two non-intersecting time intervals.

\section{Conclusions and outlook}

Let us formulate the main results obtained in the present paper. Studying the full probabilistic description of the dynamics of price increments in
two nonintersecting intraday time intervals for a set of stocks we found an approximate symmetry of the bivariate distribution ${\cal P}(x,y)$ with
respect to rotations at the multiples of $\pi/2$ and four remarkable asymmetries with respect to the axes $y=0$, $y=x$, $x=0$ and $y=-x$
characterized by the market mill patterns:
\begin{itemize}
\item{The asymmetry of the conditional response.}
\item{The asymmetry with respect to the increment's ordering.}
\item{The asymmetry of the conditional push.}
\item{The asymmetry of the mirror orderings.}
\end{itemize}

All of four market mill patterns follow from the specific structure of the general probability distribution ${\cal P}(x,y)$. The present study was
focused on describing and interpreting the market mill patterns having left many questions for future analysis \cite{LTZZ05a,LTZZ05b}. Some
particular issues we are going to address are:

\begin{itemize}
 \item{
        A detailed analysis of the shape of response pattern \cite{LTZZ05a}. Besides giving a detailed view on the structure of the response
        profile this analysis provides a direct link to a large body of econometric literature devoted to modeling of price dynamics in
        terms of regressive models such as, e.g., that in \cite{JR00}.
      }
 \item{
        We have briefly mentioned the remarkable stability of the asymmetry patterns of the individual stocks. A detailed study of these patterns
        and an analysis of the way the noisy individual patterns combine to produce a clear overall market mill structure is perhaps the most
        important problem to address \cite{LTZZ05b}.
      }
 \item{
        At the fundamental level the main challenge is to build a parsimonious model of agent strategies allowing to reproduce the specific structure of
        response asymmetry described in the present paper. In particular, it is important to discuss the issue of predictability using a more microscopic
        type of description such as used in \cite{HS94}. Work on these issues is currently in progress.
      }
\end{itemize}

The work of A.L. was partially supported by the Scientific school support grant 1936.2003.02

\section{Appendix}

Below we list stocks studied in the paper:

\medskip

 A, AA, ABS, ABT, ADI, ADM, AIG, ALTR, AMGN, AMD, AOC, APA, APOL, AV, AVP, AXP,
 BA, BBBY, BBY, BHI, BIIB, BJS, BK, BLS, BR, BSX,
 CA, CAH, CAT, CC, CCL, CCU, CIT, CL, COP, CTXS, CVS, CZN,
 DG, DE,
 EDS, EK, EOP, EXC,
 FCX, FD, FDX, FE, FISV, FITB, FRE,
 GENZ, GIS,
 HDI, HIG, HMA, HOT, HUM,
 JBL, JWN,
 INTU,
 KG, KMB, KMG,
 LH, LPX, LXK,
 MAT, MAS, MEL, MHS, MMM, MO, MVT, MX, MYG,
 NI, NKE, NTRS,
 PBG, PCAR, PFG, PGN, PNC, PX,
 RHI, ROK,
 SOV, SPG, STI, SUN,
 T, TE, TMO, TRB, TSG,
 UNP, UST,
 WHR, WY

\newpage

\begin{center}
{\bf Figure captions}
\end{center}

Figure 10. The full bivariate distribution $\log_8 {\cal P}(x,y)$, 1 min $\Longrightarrow$ 1 min.

Figure 11. The asymmetric component $\log_4 {\cal P}^{a (p)}_I (x,y)$ with respect to the axis $y=0$, 6 min $\Longrightarrow$ 6 min.

Figure 12. The asymmetric component $\log_4 {\cal P}^{a (p)}_I (x,y)$ with respect to the axis $y=0$, 30 min $\Longrightarrow$ 30 min.

Figure 13. The asymmetric component $\log_4 {\cal P}^{a (p)}_{II} (x,y)$ with respect to the axis $y=x$, 1 min $\Longrightarrow$ 1 min.

Figure 14. The asymmetric component $\log_4 {\cal P}^{a (p)}_{III} (x,y)$ with respect to the axis $x=0$, 1 min $\Longrightarrow$ 1 min.

Figure 15. The asymmetric component $\log_4 {\cal P}^{a (p)}_{IV} (x,y)$ with respect to the axis $y=-x$, 1 min $\Longrightarrow$ 1 min.

Figure 16. The asymmetric component $\log_4 {\cal P}^{a (p)}_{I} (x,y)$ with respect to the axis $y=0$, 1 min $\Longrightarrow$ 1 min, gap between
the intervals 1 min.

Figure 17. The asymmetric components $\log_4 {\cal P}^{a (p)}_I (x,y)$ with respect to the axis $y=0$, 1 min $\Longrightarrow$ 1 min, individual
stocks.

\begin{figure}[h]
 \begin{center}
 \epsfig{file=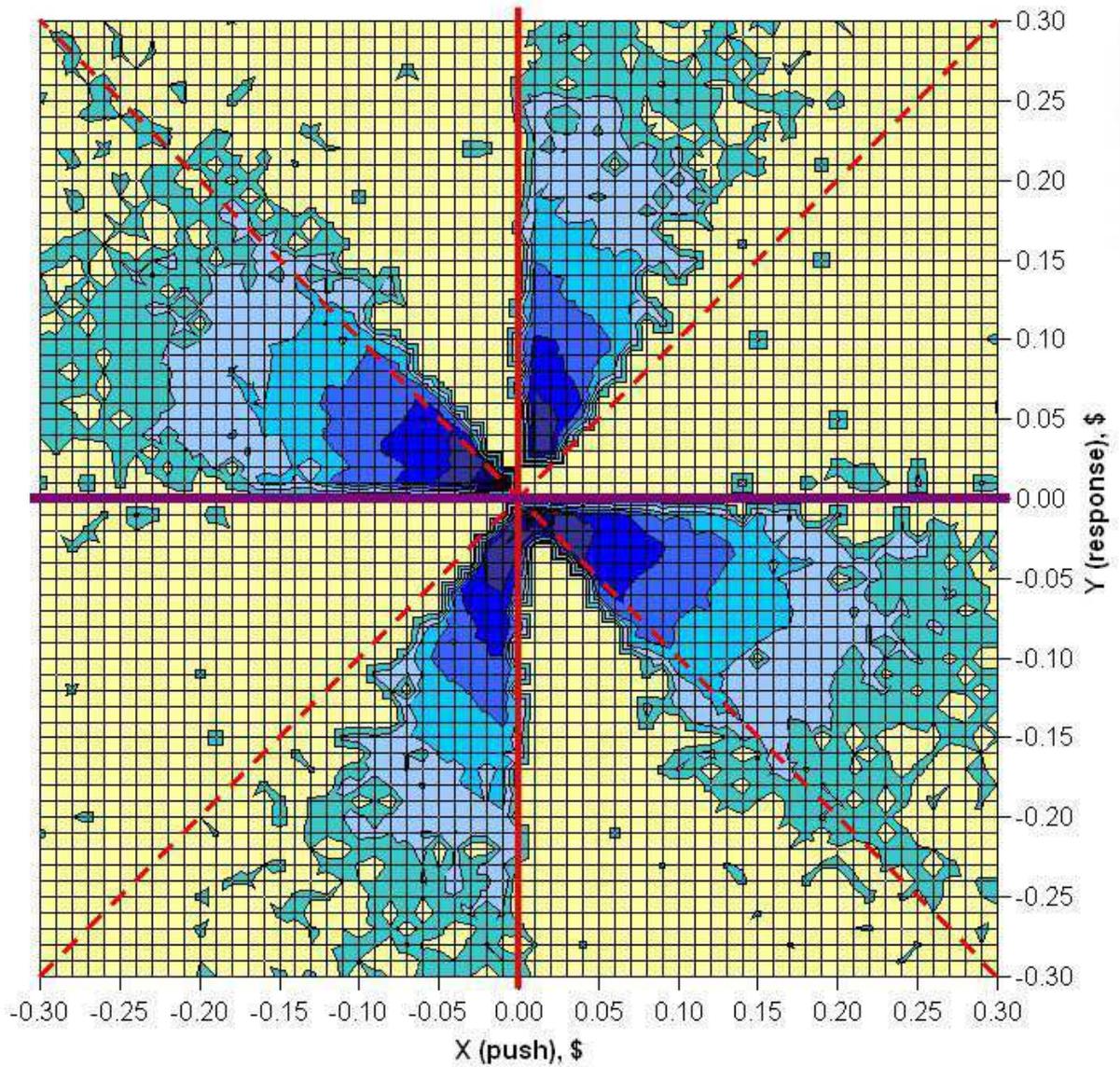,width=16cm}
 \end{center}
 \caption{Two-dimensional projection of the asymmetry of conditional distribution, 1 min $\Longrightarrow$ 1 min}
\end{figure}

\begin{figure}[h]
 \begin{center}
 \epsfig{file=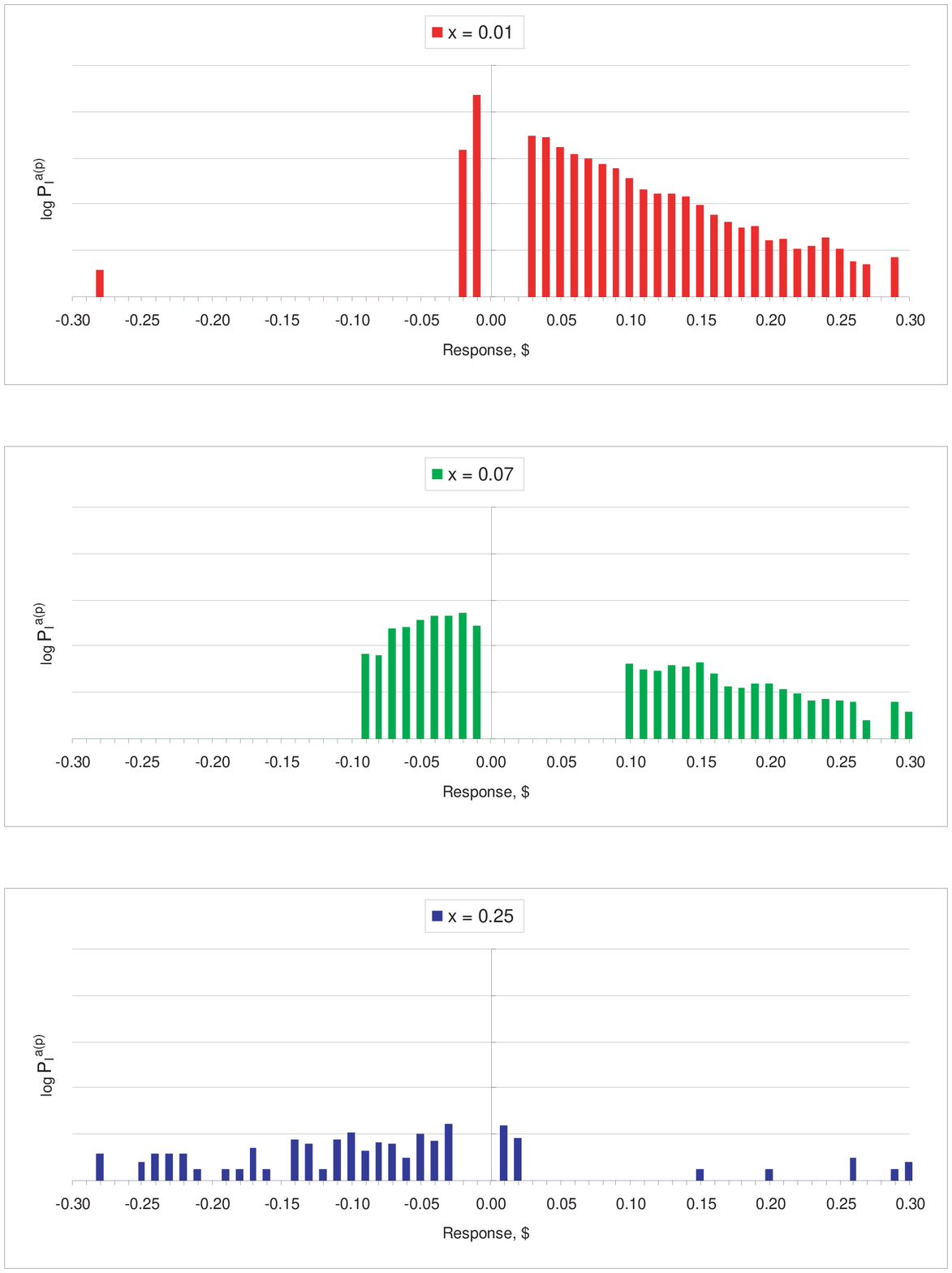,width=14cm,height=16cm}
 \end{center}
 \caption{Asymmetry of conditional distribution, 1 min $\Longrightarrow$ 1 min, selected push values $0.01, 0.07, 0.25$.}
\end{figure}


\begin{thebibliography}{99}

\bibitem{Man}
B.~Mandelbrot, "Fractal and Multifractal Finance. Crashes and Long-dependence", www.math.yale.edu/mandelbrot/webbooks/wb\_fin.html

\bibitem{Lo}
A.C.~MacKinlay, A.W.~Lo, J.Y.~Kampbell, {\it The Econometrics of Financial Markets}, Princeton, 1997; \\ A.W.~Lo, A.C.~MacKinlay, {\it A Non-Random
Walk Down Wall Sreet}, Princeton, 1999

\bibitem{SH}
A.~Shiryaev, "Essentials of Stochastic Finance: Facts, Models, Theory", World Scientific, 2003

\bibitem{BP}
J.-P.~Bouchaud, M.~Potters, {\it Theory of Financial Risk and Derivative Pricing}, Cambridge, 2000, 2003.

\bibitem{LTZ05}
A.~Leonidov, V.~Trainin, A.~Zaitsev, "On collective non-gaussian dependence patterns in high frequency financial data", ArXiv:physics/0506072,
submitted to {\it Quantitative Finance}

\bibitem{LTZZ05a}
A.~Leonidov, V.~Trainin, A.~Zaitsev, S.~Zaitsev, "Market Mill Dependence Pattern in the Stock Market: Geometry and Moments", in preparation.

\bibitem{LTZZ05b}
A.~Leonidov, V.~Trainin, A.~Zaitsev, S.~Zaitsev, "Market Mill Dependence Pattern in the Stock Market: Individual Portraits", in preparation.

\bibitem{M63}
B.~Mandelbrot, "The Variation of Certain Speculative Prices", {\it Journal\ of\ Business}\ {\bf 36} (1963), 394-419

\bibitem{MB}
B.~Mandelbrot, R.L.~Hudson, "The (Mis)behavior of Prices:  A Fractal View of Risk, Ruin, and Reward". New York: Basic Books; London: Profile Books,
2004

\bibitem{CJY05}
K.~Chen, C.~Jayprakash, B.~Yuan, "Conditional Probability as a Measure of Volatility Clustering in Financial Time Series", arXiv:physics/0503157

\bibitem{CL96}
T.F.~Crack, O.~Ledoit, "Robust Structure Without Predictability: The "Compass Rose" Pattern of the Stock Market", {\it The Journal of Finance}\ {\bf
51} (1996), 751-762

\bibitem{AV04}
A.~Antoniou, C.E.~Vorlow, "Price Clustering and Discreteness: Is there Chaos behind the Noise?", arXiv:cond-mat/0407471

\bibitem{V04}
C.E.~Vorlow, "Stock Price Clustering and Discreteness: The "Compass Rose" and Predictability", arXiv:cond-mat/0408013

\bibitem{BM03}
M.~Boguna, J.~Masoliver, "Conditional dynamics driving financial markets", ArXiv:cond-mat/0310217

\bibitem{EMS99}
P.~Embrechts, A.~McNeil, D.~Straumann, "Correlation and depenendence in risk management: properties and pitfalls", Risk Lab working paper (1999)

\bibitem{ELM01}
P.~Embrechts, P.~Lindskog, A.~McNeil, "Modelling Dependence with Copulas and Applications to Risk Management", Risk Lab working paper (2001)

\bibitem{MS01}
Y.~Malevergne, D.~Sornette, "Testing the Gaussian Copula Hypothesis for Financial Asset Dependencies", ArXix:cond-mat/0111310

\bibitem{JR01}
E.~Jondeau, M.~Rockinger, "Conditional Dependency of Financial Series: an Application of Copulas", Banque de France working paper NER 82 (2001)

\bibitem{F70}
E.F.~Fama, "Efficient Capital Markets: A Review of Theory and Empirical Work", {\it Journal of Finance}\ {\bf 25} (1970), 383-417

\bibitem{F76a}
E.F.~Fama, "Foundations of Finance", NY: Basic Books, 1976

\bibitem{F76b}
E.F.~Fama, "Efficient Capital Markets: Reply", {\it Journal of Finance}\ {\bf 31} (1976), 143-145

\bibitem{F91}
E.F.~Fama, "Efficient Capital Markets: II", {\it Journal of Finance}\ {\bf 46} (1991), 1575-1617

\bibitem{L89}
S.F.~LeRoy, "Efficient Capital Markets and Martingales", {\it Journal of Economic Literature}\ {\bf 27} (1989), 1583-1621

\bibitem{M66}
B.~Mandelbrot, "Forecasts of Future Prices, Unbiased Markets and "Martingale" Models", {\it Journal\ of\ Business}\ {\bf 39} (1966), 242-255

\bibitem{LGCMPS99}
Y.~Liu et al., "The statistical properties of the volatility of price fluctuations", {\it Phys.\ Rev.}\ {\bf E60} (1999), 1390

\bibitem{LZ03}
P.E.~Lynch, G.O.~Zumbach, "Market heterogeneities and the causal struture of volatility", {\it Quantitative Finance}\ {\bf 3} (2003), 320-331

\bibitem{LB05}
L.~Borland, J.-Ph.~Bouchaud, "On a multi-timescale statistical feedback model for volatility fluctuations", ArXiv:cond-mat/0507073

\bibitem{BDE03}
W.~Breymann, A.~Dias, P.~Embrechts, "Dependence structures for multivariate high-frequency data in finance", {\it Quantitative\, Finance}\, {\bf 3}
(2003), 1 - 14

\bibitem{JR00}
E.~Jondeau, M.~Rockinger, "Conditional volatility, skewness, and kurtosis: existence and persistence", Banque de France working paper NER 77 (2000)

\bibitem{HS94}
R.D.~Huang, H.R.~Stoll, "Market Mikrostructure and Stock Return Predictions", {\it The\ Review\ of\ Financial\ Studies}\ {\bf 7} (1994), 179-213


\end{thebibliography}
\end{document}